\documentclass[comsoc]{IEEEtran}
\usepackage{cite}
\usepackage{graphicx}
\usepackage{subcaption}
\usepackage{tikz}
\usepackage{cite}
\usepackage[T1]{fontenc} 
\usepackage{amsmath}
\usepackage[cmintegrals]{newtxmath}
\usepackage{bm}
\usepackage[fleqn]{mathtools}
\usepackage{mathtools}
\usepackage{color, colortbl}
\definecolor{Gray}{gray}{0.9}
\usepackage{makecell}
\usepackage[printwatermark]{xwatermark}
\usepackage{xcolor}

\usepackage{tabularx,ragged2e,booktabs,caption}
\usepackage{pgfplots,pgfplotstable}
\usepgfplotslibrary{statistics}

\pgfplotsset{width=6.5cm,height=5cm,compat=1.13,every axis/.append style={
		label style={font=\small},
		tick label style={font=\small}
}}
\newcommand{\squeezeup}{\vspace{-3mm}}
\newwatermark[allpages,color=gray!50,angle=90,scale=1,xpos=95,ypos=-15]{Author Copy}

\begin{document}
	\title{Optimal Slice Allocation in 5G Core Networks}
	\author{Danish Sattar and Ashraf Matrawy
		\thanks{D. Sattar, Department of Systems and Computer Engineering, Carleton University, Ottawa, ON K1S 5B6, Canada e-mail: danish.sattar@carleton.ca}
		\thanks{ A. Matrawy is with the School of Information Technology, Carleton University, Ottawa, ON K1S 5B6, Canada e-mail: ashraf.matrawy@carleton.ca}
		\thanks{This work was supported by the Natural Sciences and Engineering Research Council of Canada (NSERC) through the NSERC Discovery Grant program.}
	}%
	\maketitle
	\begin{abstract}
		5G network slicing is essential to providing flexible, scalable and on-demand solutions for the vast array of applications in 5G networks. Two key challenges of 5G network slicing are function isolation (intra-slice) and guaranteeing end-to-end delay for a slice. In this paper, we address the question of optimal allocation of a slice in 5G core networks by tackling these two challenges. We adopt and extend the work by D. Dietrich \emph{et al.}~\cite{7945385} to create a model that satisfies constraints on end-to-end delay as well as isolation between components of a slice for reliability.
		
	\end{abstract}
	\IEEEpeerreviewmaketitle
	\squeezeup
	\section{Introduction}
\begin{figure}[!ht]
		\centering
		\includegraphics[width=0.45\textwidth,keepaspectratio=true]{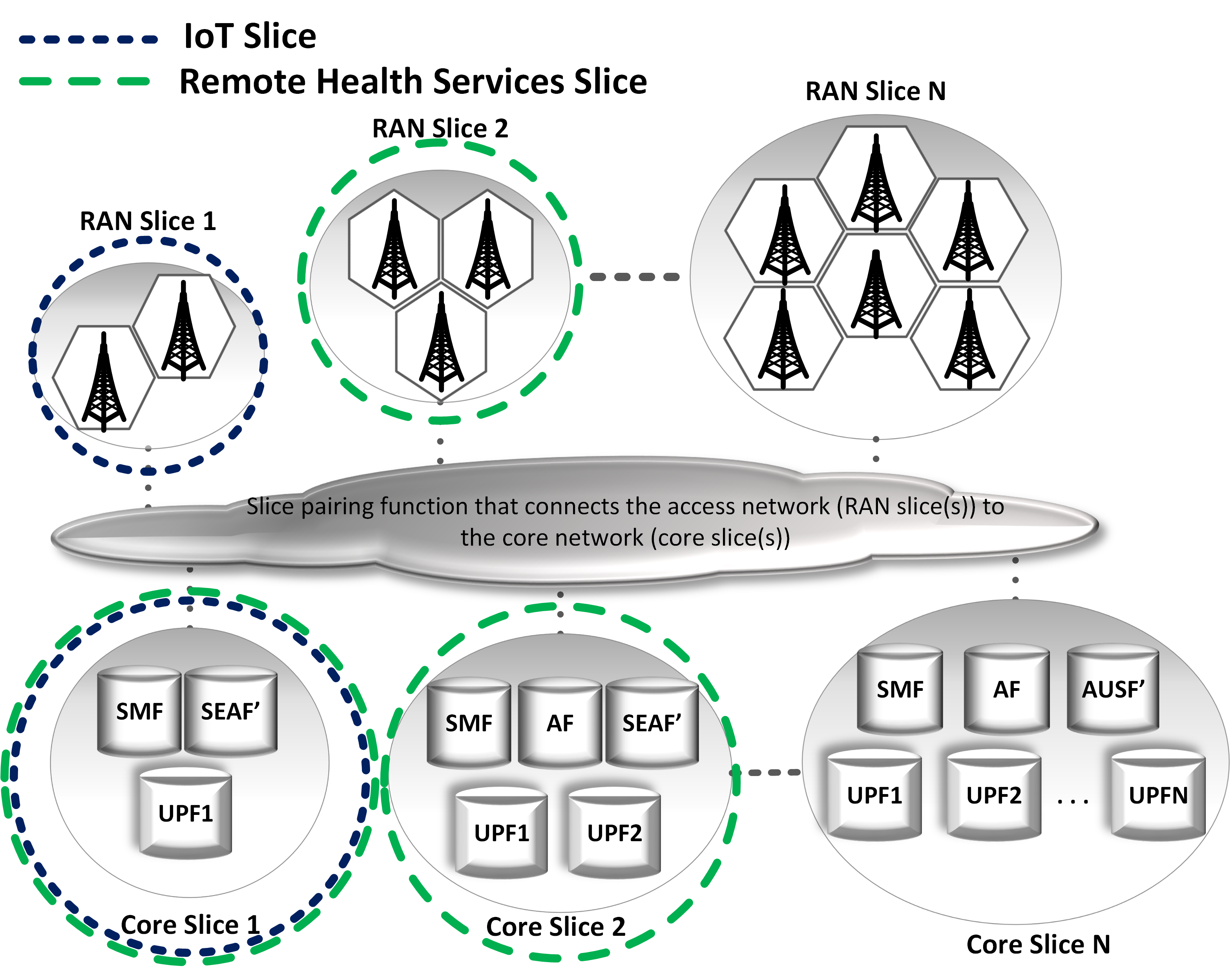}
		\caption{An example of Network Slicing - A RAN slice can be connected with
			one or more core slice and vice versa. The slice pairing function is used
			to connect the RAN-core slice pair.}\label{fig:Topology}
		\squeezeup
	\end{figure}

	The 5G network design and the standard are still in development, but it is envisioned to be an agile and elastic network. Network slicing has emerged as a key to realizing this vision.
	In 5G networks, an end-to-end network slice is a complete logical network that includes Radio Access Network (RAN) and Core Network (CN), and it has capabilities to provide different telecommunication services~\cite{rfcns}.
	An end-to-end slice is created by pairing the RAN and core network slice, but the relationship between both slices could be 1-to-1 or 1-to-M. For instance, one RAN slice could be connected to multiple core slices and vice versa~\cite{DBLP:journals/corr/LiWPU16,5gamericas}.
	Fig.~\ref{fig:Topology} shows an example of the relationship between core and RAN slices as well as 5G network slicing use cases. In Fig.~\ref{fig:Topology}, two different use cases for 5G network slicing are shown i.e. IoT, and Remote Health Services.
	
	The first issue we consider in slice allocation is intra-slice isolation (physical isolation between Virtual Network Functions (VNF) of a slice). This might be required by the slice for more reliability because if the entire slice is hosted on the same server, and if the server is compromised or becomes unavailable, the entire slice would also be affected (compromised/unavailable).
	However, if the there is some level of intra-slice isolation, the slice operator might be able to recover from partial compromise/unavailability of the network slice. We note that our aim does not include inter-slice isolation where other aspects need to be taken into consideration including but not limited to physical isolation, hardware-based isolation, virtual machine based isolation~\cite{8104638}. However, our focus is on providing on-demand physical isolation between different VNFs of a slice for added reliability and security.
	
	The second issue we consider is the end-to-end delay. 5G networks have strict requirements for the end-to-end delay. To support real-time applications (e.g., health services, autonomous driving, etc.) 5G network needs to guarantee end-to-end delay for a certain application across the network (only considering end-to-end delay for a core network slice).
	
	In this paper, we address the question of optimal slice allocation in 5G core networks (virtual Evolved Core (vEPC)). We do this by adopting and extending VNF placement in the LTE core network presented by D. Dietrich \emph{et al.}~\cite{7945385}. Our \textbf{contributions} are to (1) guarantee end-to-end delay, (2) provide intra-slice isolation for slice allocation and (3) find a minimum delay path between the slice components. We aim to provide an optimal solution for allocating a core network slice in 5G networks. The formulation we use for the optimization model is Mixed-Linear Integer Programming (MILP). We take into consideration some of the core requirements for allocating a 5G network slice. We consider the physical isolation requirement between different components of a slice for a variable degree of reliability. The optimization model also ensures the end-to-end delay required by a core network slice. A 5G network slice creation would be dynamic, and a slice could have a variable number of components that require on-demand service chaining (network slices might have different combination of VNFs). For instance, a slice could have several components e.g., Authentication Server Function (AUSF), Security Anchor Function (SEAF), Session Management Function (SMF), Application Function (AF) and several User Plane Functions (UPF) with on-demand service chaining between them. We are aware that there are several other requirements and properties that need to be addressed before a complete end-to-end 5G slice can be instantiated but those requirements are out-of-the-scope of this work.%
	
	\section{Related Work}
	\label{sec:relatedwork}
	
	V. Sciancalepore et al. \cite{2018arXiv180103484S} have proposed a practical implementation of network slicing. The proposed model aims to provide an efficient network slicing solution by analyzing the past network slicing information.
	D. Dietrich \emph{et al.}~\cite{7945385} proposed linear programming formulation for the placement of VNFs in the LTE core network. In the proposed algorithm, they provided a balance between optimality and time complexity. R. Ford \emph{et al.}~\cite{DBLP:journals/corr/FordSMJR17} proposed optimal VNF placement for the SDN-based 5G mobile-edge cloud. Their optimization algorithm provides resilience by placing VNFs in distributed data centers. A. Baumgartner et al. \cite{7116162} have presented optimal VNF placement for the mobile virtual core. They used the cost of placement to allocate the VNFs. In their problem formulation, they considered physical network constraints for storage, processing, and switching capacity as well as service chains when allocating VNFs to the physical substrate network.
	S. Agarwal et al. \cite{Francesco2017JointVP} used a queuing model to perform VNF placement in 5G networks. Latency was used as the primary Key Performance Indicator (KPI) to formulate the optimization problem.%
	
	\section{MILP Formulation}
	\label{sec:mathmodel}
	In this section, we will explain the optimization model we used in this paper. We are adopting and extending the work presented by D. Dietrich \emph{et al.} in~\cite{7945385}. The focus of their work was on the LTE cellular core and placement of network functions in an optimal manner while load balancing the resources. They transformed the optimization problem into Linear Programming (LP) problem by relaxing some MILP constraints to reduce time complexity. We adopt their model to achieve optimal slice allocation. Our objective is to allocate 5G core network slice VNFs optimally to provide intra-slice isolation for added reliability. We also fulfill one of the core 5G network requirements by guaranteeing the end-to-end core slice delay.
	
	In the following MILP formulation, we use the network model and variables from~\cite{7945385}. In that model, each request is associated with a computing demand ($g^i$) and bandwidth requirement ($g^{ij}$).  Additionally, for our slice request, we consider end-to-end delay ($d_{E2E}$) and intra-slice isolation (reliability) required between the VNFs ($K_{rel}$). We use the following objective function.
	
	\begin{equation}\label{eq1}
	\begin{multlined}
	$Minimize$\\
	\sum _{i\in{V}_{F}}\sum _{u\in {V}_{S}}\left( 1- {\frac{{r}_{u}}{r_{u,max}}}\right) g^i x^i_u \gamma^i_u \\
	+  \sum _{(i,j)\in{E}_{F}}\sum _{\substack{(u,v)\in {E}_{S} \\(u\neq v)}} L_{uv} f^{ij}_{uv}
	\end{multlined}
	\end{equation}
	
	subject to:
	\begin{align}\label{eq9}
		\begin{multlined}
			\sum _{i \in V_F} x^i_u\leq K_{rel} \hspace{.5cm}\forall u \in V_S, K_{rel}=1,2,3...
		\end{multlined}
	\end{align}
	
	\begin{equation}\label{eq6}
	\begin{multlined}
	\sum _{\substack{(i,j)\in {E}_{F} }} \sum _{\substack{(u,v)\in {E}_{S}\\u\neq v}}\left(\dfrac{f^{ij}_{uv}}{g^{ij}} L_{uv}\right) + \sum_{i\in V_F} \alpha^i\leq d_{E2E}
	\end{multlined}
	\end{equation}
	
	\begin{equation}\label{eq7}
	\begin{multlined}
	\sum _{i \in V_F} g^i \leq \sum_{u \in V_S} r_u
	\end{multlined}
	\end{equation}
	
	\begin{equation}\label{eq8}
	\begin{multlined}
	\sum _{(i,j) \in E_F} g^{ij} \leq \sum_{(u,v) \in E_S} r_{uv}
	\end{multlined}
	\end{equation}
	
The objective function (\ref{eq1}) will assign the incoming slice requests to the least utilized server and find a path with minimum delay. The first term is identical to the objective function in~\cite{7945385} while the second one differs in the way we select paths between VNFs. The first term of the objective function assigns computing demands to the least utilized physical servers. The parameter $\gamma_u^i$ used to avoid infeasible mapping of the VNF/server combination. The second term takes into consideration the physical link delay ($L_{uv}$). Each time when a virtual link $(i,j)\in E_F$ is assigned to a physical link $(u,v)\in E_s$, it increases $L_{uv}$. $L_{uv}$ is a function of link utilization, and it is calculated using eq. (\ref{eq10}), where $L_{uv,init}$ is the initial delay assigned to the link $(u,v)\in E_s$. Minimizing both terms will result in the assignment of a network slice to the least utilized servers, and it will find a path with least delay between the slice components (D. Dietrich \emph{et al.}~\cite{7945385} did not consider the minimum delay path).
	\begin{equation}\label{eq10}
	\begin{multlined}
	L_{uv}=(1-\frac{r_{uv}}{r_{uv,max}})\:2.5\:ms + L_{uv,init}\hspace{.3cm} \forall (u,v) \in E_S
	\end{multlined}
	\end{equation}
	
The objective function is subjected to several MILP constraints that we will explain next. In our work, in addition to the constraints listed here, we use constraints (2-5) and (9-10) from~\cite{7945385}. We are not listing/describing all the parameters and constraints due to the space limitation. If the slice has requested that each VNF needs to be assigned to different physical servers, constraint (\ref{eq9}) will provide the desired degree of reliability (intra-slice isolation) for the slice ($K_{rel}$). The end-to-end delay for the 5G network is an important requirement.
Constraint (\ref{eq6}) enforces the end-to-end delay requirement for the core network slice\footnote{We note that meeting end-to-end delay requirements would need traffic engineering which is outside the scope of this paper.}. It includes the delay incurred along the entire path and the processing delay of each VNF ($\alpha^i$). Since the partial or incomplete assignment of the slice components serves no purpose, constraints (\ref{eq7}) and (\ref{eq8}) ensure that the remaining computing and bandwidth capacity of the entire data center is enough to accommodate the slice creation request. $x_u^i \in {0,1}$ and $f_{uv}^{ij} \geq 0$ are binary and real variables, respectively.
	
	\section{Results and Discussion}
	\label{sec:resutls}
	\begin{table}[htbp]
		\centering
		\captionof{table}{Simulation parameters} \label{tab:parameters}
		\begin{tabular}{r||l}\toprule[1.5pt]
			
			\bf Parameter						& \bf Value 	\\\midrule
			\rowcolor{Gray}
			CPU capacity/server ($r_{u,max}$)	&	12.0 GHz	\\
			Total Servers ($V_S$)				&   200			\\
			\rowcolor{Gray}
			Total Slice Requests 				& 	200 		\\
			$K_{rel}$ 							&	1-10		\\
			\rowcolor{Gray}
			VNF/slice ($V_F$)					&	10			\\
			Bandwidth request/slice ($g^{ij}$)  &   30-70 Mbps	\\
			\rowcolor{Gray}
			VNF CPU request/slice ($g^i$)       &   0.5-2.0 GHz	\\
			\rowcolor{Gray}
			$\alpha^i$                          &   0.3-2.0 ms	\\
			$\epsilon$                          &   $10^{-10}$	\\
			\bottomrule[1.25pt]
			\end {tabular}\par
			\bigskip
		\end{table}
		\pgfplotstableread[row sep=\\,col sep=&]{
			K	& CPU  & BW 	& Requests 	& solver\\
			1   & 50.67 & 32.76 	& 48.0 		& 138.86\\
			2   & 91.82 & 31.97 	& 87.5 		& 2764.1\\
			3   & 96.81 & 29.40 	& 93.0 		& 1446.7\\
			4   & 97.35 & 23.52 	& 92.5 		& 62.2\\
			5  	& 96.12 & 15.68		& 93.0 		& 6.54\\
			6  	& 96.33 & 15.72		& 93.0 		& 2.29\\
			7  	& 96.39 & 16.81 	& 93.0 		& 1.85\\
			8  	& 96.43 & 16.99 	& 93.0 		& 1.5\\
			9  	& 96.99 & 16.35 	& 93.0 		& 1.21\\
			10  & 96.93 & 13.70 	& 93.5		& 0.85\\
		}\mydata
		
		\pgfplotstableread[row sep=\\,col sep=&]{
			K	& CPU  & BW 	& Requests 	& solver\\
			1   & 20.21 & 15.47 	& 19.5 		& 3.71\\
			2   & 41.94 & 15.28 	& 39.5 		& 255.36\\
			3   & 50.47 & 14.98 	& 47.5 		& 16.59\\
			4   & 63.02 & 14.55 	& 59.0 		& 11.14\\
			5  	& 81.19 & 14.98		& 77.0 		& 0.43\\
			6  	& 82.06 & 15.95		& 78.0 		& 0.44\\
			7  	& 80.63 & 15.71 	& 76.5 		& 0.43\\
			8  	& 78.50 & 15.06 	& 74.5 		& 0.43\\
			9  	& 78.09 & 13.67 	& 74.0 		& 0.46\\
			10  & 88.69 & 13.69 	& 85.5 		& 0.22\\
		}\mydataBW
		
		To test the optimization model, we used MATLAB to simulate 5G core network and slice requests. AMPL is used to model optimization algorithm and CPLEX 12.8.0.0 is used as MILP solver. The optimization algorithm is evaluated on Intel Core i7 3.2 GHz with 32 GB RAM.
		\begin{figure}[!ht]
			\centering
			\includegraphics[width=0.5\textwidth,keepaspectratio=true]{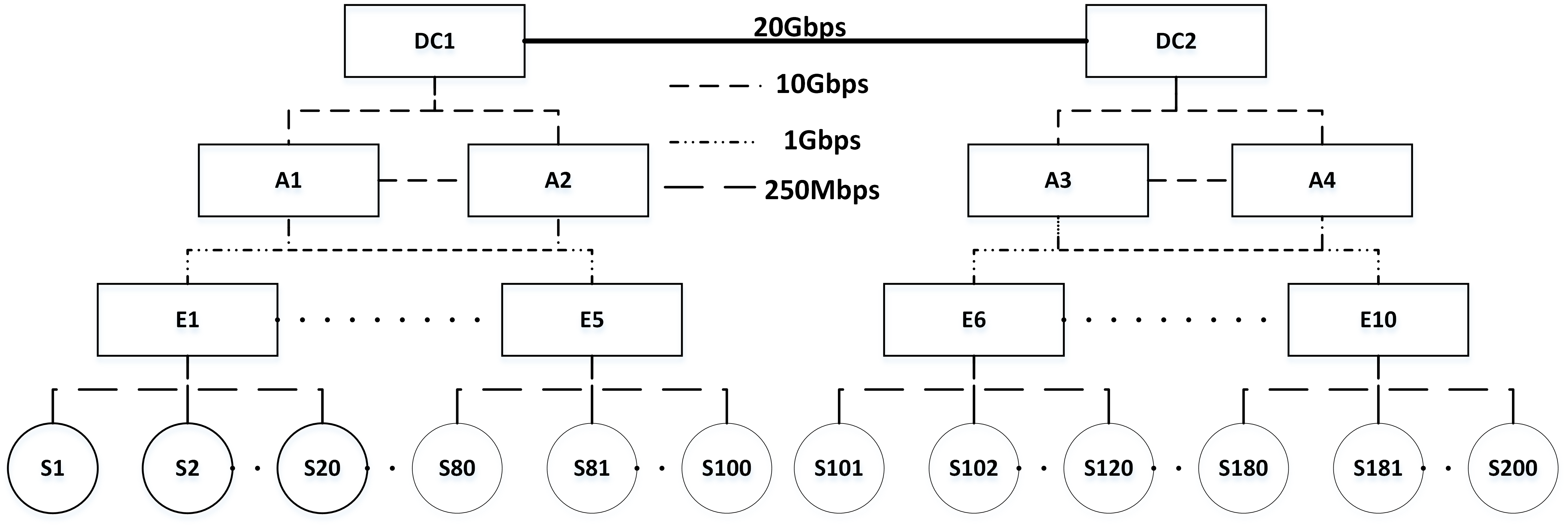}
			\caption{Simulation topology. S1-S200, E1-E10, A1-A4, and DC1-DC2 represent
				physical servers, Edge, Aggregation and Datacenter switches,
				respectively}\label{fig:Topology1}
		\end{figure}%
		
We simulate 200 physical servers that can host different types of VNFs. Other parameters used for the evaluation are listed in Table~\ref{tab:parameters}. In our simulations, we vary the level of intra-slice isolation using the $K_{rel}$ parameter. This parameter provides the upper limit for how many VNFs can be placed on one physical server. The model guarantees the requested computing resources, bandwidth resources, and end-to-end delay for a slice in the current network state. After allocating each slice, we update the remaining computing and bandwidth resources.  The flow link delay $L_{uv}$ can be dynamic. For instance, when the network is congested, this parameter can be updated to reflect the current state of the network, but we did not consider this case.

In our simulations, we used two configurations for link bandwidth (Servers$\leftrightarrow$EdgeSwitches). In the first configuration as shown in fig.~\ref{fig:Topology1}, the link bandwidth between servers and edge switches is set to $250$ Mbps. In this case, the overall system performance is limited by the available CPU capacity (CPU bound). Therefore our simulated slice requests, the CPU capacity of the physical servers becomes the limiting factor when allocation slices rather than the link bandwidth. In the second configuration, the link bandwidth (Servers$\leftrightarrow$EdgeSwitches) is set to $100$ Mbps (Bandwidth bound). In this case, the overall system performance is limited by the available link bandwidth between servers and Edge switches.
Please note that in all the presented results, the simulation setup was "CPU bound" unless otherwise stated.%
		\subsection{Intra-slice isolation}
		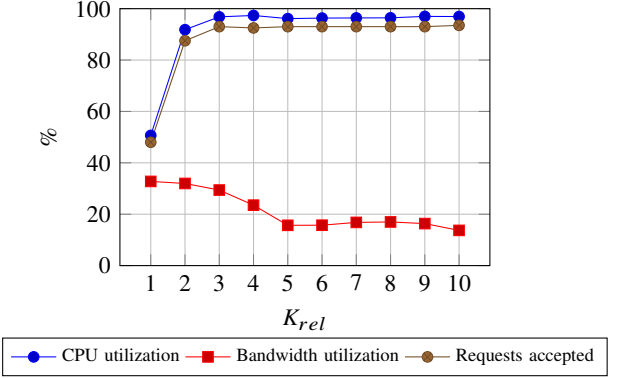
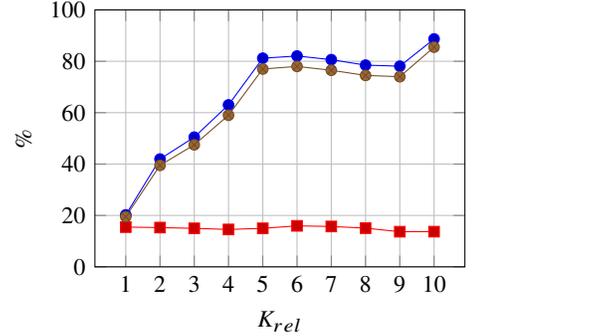
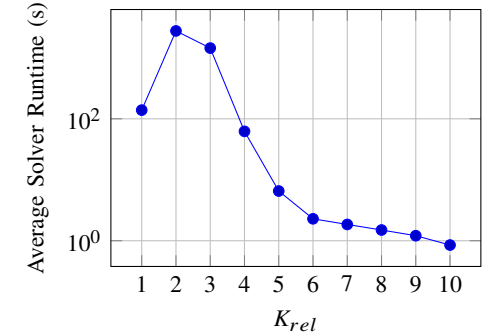
\begin{figure}[ht!]
			\centering
			\begin{subfigure}[t]{0.4\textwidth}
				\centering
				\begin{tikzpicture}[baseline]
				\begin{axis}[
				ymin=0,  ymax=100,
				grid=major,
				ylabel={\%},
				xlabel={$K_{rel}$},
				legend style={font=\scriptsize,at={(0.5,-0.28)},
					anchor=north,legend columns=-1},
				symbolic x coords={1,2,3,4,5,6,7,8,9,10},
				xtick=data,
				]
				\addplot table[x=K,y=CPU]{\mydata};
				\addplot table[x=K,y=BW]{\mydata};
				\addplot table[x=K,y=Requests]{\mydata};
				\legend{CPU utilization, Bandwidth utilization, Requests accepted}
				\end{axis}
				\end{tikzpicture}
				\caption{CPU utilization, bandwidth utilization and requests accepted
					for varying levels of $K_{rel}$ (CPU Bound)}
				\label{fig:utilization}
			\end{subfigure}
			
			\begin{subfigure}[t]{0.4\textwidth}
				\centering
				\begin{tikzpicture}[baseline]
				\begin{axis}[
				ymin=0,  ymax=100,
				grid=major,
				ylabel={\%},
				xlabel={$K_{rel}$},
				legend style={font=\scriptsize,at={(0.5,-0.28)},
					anchor=north,legend columns=-1},
				symbolic x coords={1,2,3,4,5,6,7,8,9,10},
				xtick=data,
				]
				\addplot table[x=K,y=CPU]{\mydataBW};
				\addplot table[x=K,y=BW]{\mydataBW};
				\addplot table[x=K,y=Requests]{\mydataBW};
				\legend{CPU utilization, Bandwidth utilization, Requests accepted}
				\end{axis}
				\end{tikzpicture}
				\caption{CPU utilization, bandwidth utilization and requests accepted
					for varying levels of $K_{rel}$ (Bandwidth Bound)}
				\label{fig:utilizationBW}
			\end{subfigure}
			
			\begin{subfigure}[t]{0.4\textwidth}
				\centering
				\begin{tikzpicture}[baseline]
				\begin{axis}[
				grid=major,
				ylabel = {Average Solver Runtime (s)},
				ymode = log,
				xtick={1,2,3,4,5,6,7,8,9,10}, xticklabels={1,2,3,4,5,6,7,8,9,10},
				xlabel={$K_{rel}$},
				]
				\addplot table[x=K,y=solver]{\mydata};
				\end{axis}
				\end{tikzpicture}
				\caption{Average solver runtime (seconds) for varying levels of $K_{rel}$} \label{fig:sisolation}
			\end{subfigure}
			\caption{Intra-slice isolation simulation results}
		\end{figure}
		
		In the first part of the simulation, we fix the end-to-end delay ($d_{E2E}$) to a relatively high value ($500$ ms) to minimize its affect on the results and vary the  levels of intra-slice isolation ($K_{rel}$). Fig.~\ref{fig:utilization} shows the overall average system utilization for CPU and bandwidth resources and accepted requests for different levels of intra-slice isolation. The system is CPU bound, so that overall system bandwidth is higher than total requested bandwidth, hence we see relatively low bandwidth utilization.  When slices request intra-slice isolation where $K_{rel}<4$, the bandwidth utilization is higher because all VNFs would have to utilize physical links to communicate with each other. Whereas, when we relax the intra-slice isolation requirement, we get lower network utilization (i.e., $K_{rel}\ge4$). The reason is that as we can allocate more VNFs on the same physical server and the communication between the VNFs does not
		involve physical communication link, we see lower network activity in this case. However, there is a marginal difference in CPU utilization and requests accepted for variable levels of $K_{rel}\ge2$.
		
		We also simulated another topology where the system was bandwidth bound. Fig.~\ref{fig:utilizationBW} shows the overall system utilization for CPU, bandwidth and requests accepted. The performance of the system is worse compared to when the system is CPU bound (Fig.~\ref{fig:utilization}).
		
		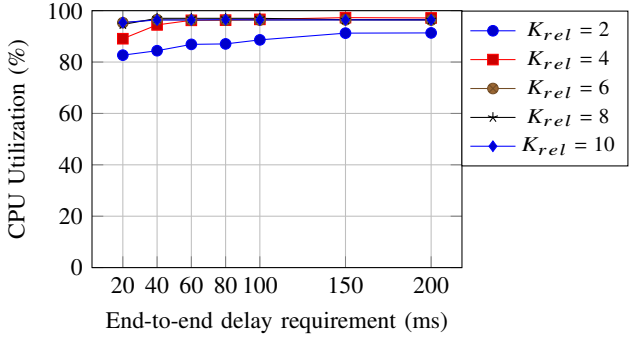
\begin{figure}[t!]
			\centering
			\begin{tikzpicture}[baseline]
			\begin{axis}[
			ymin=0,  ymax=100,
			grid=major,
			ylabel={CPU Utilization (\%)},
			xlabel={End-to-end delay requirement (ms)},
			legend style={font=\footnotesize,at={(1.45,0.697)},
				anchor=east,legend columns=1},
			xtick={20,40,60,80,100,150,200},
			]
			\addplot table[x=E2E,y=K2] {E2ECPU.txt};
			\addplot table[x=E2E,y=K4] {E2ECPU.txt};
			\addplot table[x=E2E,y=K6] {E2ECPU.txt};
			\addplot table[x=E2E,y=K8] {E2ECPU.txt};
			\addplot table[x=E2E,y=K10] {E2ECPU.txt};
			\legend{$K_{rel}=2$,$K_{rel}=4$,$K_{rel}=6$,$K_{rel}=8$,$K_{rel}=10$}
			\end{axis}
			\end{tikzpicture}
			\caption{CPU utilization for different end-to-end delay requirements} \label{fig:e2ecpu}
		\end{figure}
		Fig.~\ref{fig:sisolation} provide some interesting results for the average solver runtime. Obviously, with stricter requirements for intra-slice isolation are, more time is required to find an optimal solution for allocation of slice components and to find an optimal path with least delay. A factor that impacts these values is that when the requirement for intra-isolation are flexible, the optimization algorithm can place more components on the same physical system and it would eliminate the need to find optimal paths between these components. We can see this behavior when $K_{rel}>4$ in Fig.~\ref{fig:sisolation}. However, as we can see, when a slice requests that no more than two or three VNFs can be placed on a single physical server, there is a significant variation in solver runtime. We ran these simulations multiple times and using multiple parameter value and each time we obtain almost identical results. We have not been able to identify the reason behind the anomalous behavior for $K_{rel}=2$ and $K_{rel}=3$. 
		\subsection{End-to-end delay}
		In the second part of the simulation, we use different end-to-end delay requirements. Please note that we ran simulations for $K_{rel}=1$ to $K_{rel}=10$ but results are only shown for a few values of $K_{rel}$ to present more readable graphs.
		The end-to-end delay parameter has a noticeable effect on CPU utilization because setting $K_{rel}\le2$ reduces the number of available solutions as shown in Fig. \ref{fig:e2ecpu}. However, this effect becomes minimal when $d_{E2E}\ge150$. We note that the CPU utilization shows the same behaviour as the request acceptance rate (not shown here).
		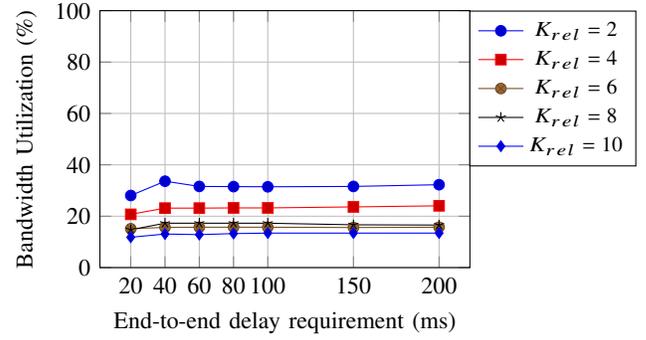
\begin{figure}[t!]
			\centering
			\begin{tikzpicture}[baseline]
			\begin{axis}[
			ymin=0,  ymax=100,
			grid=major,
			ylabel={Bandwidth Utilization (\%)},
			xlabel={End-to-end delay requirement (ms)},
			legend style={font=\footnotesize,at={(1.45,0.697)},
				anchor=east,legend columns=1},
			xtick={20,40,60,80,100,150,200},
			]
			\addplot table[x=E2E,y=K2] {E2EBW.txt};
			\addplot table[x=E2E,y=K4] {E2EBW.txt};
			\addplot table[x=E2E,y=K6] {E2EBW.txt};
			\addplot table[x=E2E,y=K8] {E2EBW.txt};
			\addplot table[x=E2E,y=K10] {E2EBW.txt};
			\legend{$K_{rel}=2$,$K_{rel}=4$,$K_{rel}=6$,$K_{rel}=8$,$K_{rel}=10$}
			\end{axis}
			\end{tikzpicture}
			\caption{Bandwidth utilization for different end-to-end delay requirements} \label{fig:e2ebw}
			
		\end{figure}
		
		\begin{figure}[t!]
			\centering
			\begin{tikzpicture}[baseline]
			\begin{axis}[
			ymode=log,
			grid=major,
			ylabel={Average Solver Runtime (s)},
			xlabel={End-to-end delay requirement (ms)},
			legend style={font=\scriptsize,at={(1.427,0.71)},
				anchor=east,legend columns=1},
			xtick={20,40,60,80,100,150,200},
			]
			\addplot+ table[y=K2] {E2ETime.txt};
			\addplot+ table[y=K4] {E2ETime.txt};
			\addplot+ table[y=K6] {E2ETime.txt};
			\addplot+ table[y=K8] {E2ETime.txt};
			\addplot+ table[y=K10] {E2ETime.txt};
			\legend{$K_{rel}=2$,$K_{rel}=4$,$K_{rel}=6$,$K_{rel}=8$,$K_{rel}=10$}
			\end{axis}
			\end{tikzpicture}
			\caption{Average solver runtime (seconds) for different end-to-end delay requirements} \label{fig:e2esrt}
		\end{figure}
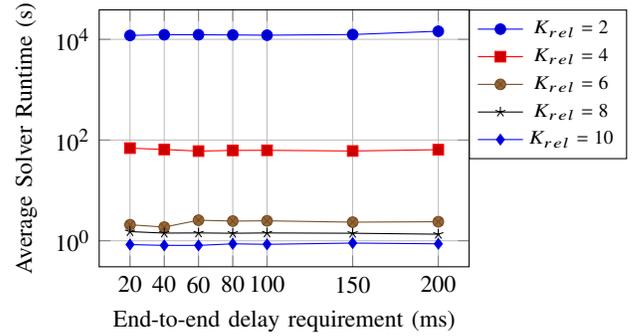
		
		Fig. \ref{fig:e2ebw} shows that different end-to-end delay requirements have minimal impact on overall bandwidth utilization for all levels of $K_{rel}$. Fig.~\ref{fig:e2esrt} shows the average solver runtime for different end-to-end delay requirements. We can see a consistent behavior for all levels of intra-slice isolation.
		
		\section{Conclusion}
		\label{sec:con}
		In this paper, we addressed the optimal allocation of 5G core network slices. The optimization model provides intra-slice isolation as well as ensures that the end-to-end delay meets the minimum requirement. We evaluated the optimization model by simulating a virtualized mobile core. Our evaluation shows that when there is little or no restriction on the intra-slice isolation ($K_{rel} > 2$), CPU utilization is increased and the demand for bandwidth is reduced due to the reduction between inter-machine communications. On the other hand, stricter intra-slice isolation ($K_{rel} \leq 2$) requires more bandwidth and leads to relatively lower CPU utilization.
		\bibliographystyle{IEEEtran}
		\bibliography{5G_bib_cletter}
	\end{document}